# What can one learn from two-state single molecule trajectories?

Ophir Flomenbom,*[†] Joseph Klafter,* and Attila Szabo[†]

* *School of Chemistry, Raymond & Beverly Sackler Faculty of Exact Sciences, Tel Aviv University, Ramat Aviv, Tel Aviv 69978, Israel*;

[†] *Laboratory of Chemical Physics, National Institute of Diabetes and Digestive and Kidney Diseases, National Institutes of Health, Bethesda, Maryland 20892*;

## ABSTRACT

**A time trajectory of an observable that fluctuates between two values (say, *on* and *off*), stemming from some unknown multi-substate kinetic scheme, is the output of many single molecule experiments. Here we show that when all successive waiting times along the trajectory are uncorrelated the *on* and the *off* waiting time probability density functions (PDFs) contain all the information. By relating the lack of correlation in the trajectory to the topology of kinetic schemes, we can immediately specify those kinetic schemes that are equally consistent with experiment, which means that it is impossible to differentiate between them by any sophisticated analyses of the trajectory. Correlated trajectories, however, contain additional information about the underlying kinetic scheme, and we consider the strategy that one should use to extract it. An example is given on correlations in the activity of individual lipase molecules.**

# INTRODUCTION

Since the first patch clamp measurements (Neher and Sakmann, 1976), great advances have been made in our ability to look at complex systems on the single molecule level (Moerner and Orrit, 1999; Weiss, 1999; Nie et al., 1994; Shera et al., 1990; Mets et al., 1994; Ha et al., 1999; Schuler et al., 2002; Yang et al., 2003; Rhoades et al., 2003; Wennmalm et al., 1997; Bokinsky et al., 2003; Lu et al., 1998; Edman et al., 1999; Edman and Rigler, 2000; Velonia et al.; Flomenbom et al.; Kasianowicz et al., 1996). In an important class of such experiments, the output is a time-series (trajectory) of *on-off* events (Figs. 1A & 1B). For example, in patch clamp measurements (Neher and Sakmann, 1976), one records the ion current through a membrane-pore under an applied electric field for a long time. The fluctuations between two values of the current are attributed to conformational changes that result in opening and closing the membrane-pore. From the two-state current trajectory, one wishes to learn about the dynamics of conformational changes of the membrane-pore. In single enzyme activity measurements (Lu et al., 1998; Edman et al., 1999; Edman and Rigler, 2000; Velonia et al.; Flomenbom et al.), one monitors photon counts as a function of time, which are then collected into bins giving rise to the trajectory. A two-state trajectory is obtained when either the enzyme itself switches between a fluorescent state and a non-fluorescent state (Lu et al., 1998), or a non-fluorescent substrate is transformed into a fluorescent product (Edman et al., 1999; Edman and Rigler, 2000; Velonia et al.; Flomenbom et al.). By studying this system, one wishes to deduce the mechanism of the enzymatic activity.

In practice, noise induced fluctuations in the signal occur around the *on* and the *off* values. The ability to restore reliably the noiseless trajectory from the experimental output (i. e. to deconvolute the noise) depends roughly on the difference between these values relative to the sum of the amplitudes of the noise in each of the states. Here, we assume that we are given a noiseless two-state trajectory.

Such a two-state trajectory contains information about the underlying mechanism, which we describe by a kinetic scheme in which each substate belongs either to the *on* state or to the *off* state. The kinetic scheme may have a large number of substates (Figs. 2A-2F), and a net flow at steady-state along some of the connections (Figs. 2G-2I) (i. e. a non-equilibrium steady state), when an external source of energy is present (Hill, 1985). The goal of single molecule measurements that result in two-state trajectories is to learn as much as possible about the underlying kinetic scheme.

## FROM TRAJECTORIES TO KINETIC SCHEMES

The basic functions that are easily obtained from single molecule two-state time-series are the waiting time probability density functions (PDFs) of the *on* state, $\phi_{on}(t)$, and of the *off* state, $\phi_{off}(t)$. These functions, which cannot be found from bulk experiments, can be calculated for any kinetic scheme (Cao, 2000). Clearly, any proposed kinetic scheme must reproduce $\phi_{on}(t)$ and $\phi_{off}(t)$. However, when $\phi_{on}(t)$ and $\phi_{off}(t)$ are multi-exponentials, several models will fulfill this requirement, and their number increases with the complexity of the waiting time PDFs (the trajectories on Figs. 1A & 1B have the same waiting time PDFs, but were produced from different kinetic schemes). Can one discriminate between kinetic schemes that lead to the same $\phi_{on}(t)$ and $\phi_{off}(t)$ by looking at the trajectory in more detail?

A trajectory is completely described by $\phi_{on}(t)$ and $\phi_{off}(t)$ only when waiting times along the trajectory are uncorrelated. Therefore, kinetic schemes that lead to uncorrelated trajectories with the same $\phi_{on}(t)$ and $\phi_{off}(t)$ cannot be distinguished by the trajectory analysis. This means that the trajectory from such a kinetic scheme does not contain information about the connectivity of substates within the two states, which, as shown below, is a consequence of a specific connectivity between substates of different states. We say that such schemes are *reducible* to a two-state-semi-Markovian (TSSM) scheme (Fig. 2J). A TSSM process is one where the *on* [*off*] waiting times are drawn randomly and independently out of a non-exponential $\phi_{on}(t)$ [$\phi_{off}(t)$]. In the literature, the term non-Markovian is often used for any process with non-exponential waiting time PDFs. However, here we reserve this term to describe a trajectory of correlated waiting times.

The most straightforward test for correlation in the trajectory is based on the two successive waiting times PDFs, $\phi_{x,y}(t_1,t_2)$, $x,y = on, off$. A trajectory shows no correlations when $\phi_{x,y}(t_1,t_2)$ can be written, for every $x$ and $y$, as a product of the individual waiting time PDFs, $\phi_x(t_1)$ and $\phi_y(t_2)$,

$$\phi_{x,y}(t_1,t_2) = \phi_x(t_1)\phi_y(t_2) \qquad ; \qquad x,y = on, off . \qquad (1)$$

It is sufficient to demand only the factorization of the two successive waiting times PDFs because there are only two observable states, so when all two successive waiting times PDFs are factorized, higher order successive waiting times PDFs, e. g. $\phi_{x,y,z}(t_1,t_2,t_3)$ $x,y,z = on, off$, will also be factorized. Since higher order successive waiting times PDFs determine all the statistical properties of the trajectory and these factorize when Eq. (1) is fulfilled, it follows that for uncorrelated trajectories $\phi_{on}(t)$ and $\phi_{off}(t)$ contain all the information in the time-series.

Kinetic schemes are reducible [i. e. fulfill Eq. (1)] regardless of the system parameters if and only if after every transition from the *on* state to the *off* state, the *off* substates are populated with the same initial probabilities, and vice versa. This occurs only for a very specific connectivity between the *on* and the *off* substates of schemes, and we now give a full characterization of the reducible schemes. When only reversible connections between substates are present, a scheme is reducible when the *on* and the *off* regions are connected through one substate (Figs. 2A-2F), called a gateway substate. In general, there are two types of gateway substates. A type 1 gateway substate is one where all the transitions from the other state *enter* it (the *on* substate 1 on Fig. 2G). A type 2 gateway substate is one where all the transitions to the other state *originate* from it (the *on* substate 2 on Fig. 2G). Thus, for a reducible scheme with only reversible connections, the gateway substate is of both types simultaneously. For a kinetic scheme with a non-equilibrium steady state, there are three combinations of gateway substates that lead to a reducible scheme: (a) two gateway substates of different types in the same state (Fig. 2G), and (b) & (c) two gateway substates of the same type, either type 1 (Fig. 2H) or type 2 (Fig. 2I), in different states. Note that the above requirements are the minimal ones and a reducible scheme can possess more than two gateway substates. Since our argument relies only on the connectivity of the scheme, the reducible schemes can be characterized by any substate waiting time PDFs and not just the Markovian (exponential) one. Additionally, other less general schemes can fulfill Eq. (1), thus are reducible, because of symmetry for special choices of the transition rates.

As an example, consider the two schemes shown in Fig. 2B and Fig. 2C, each containing *n off* substates and one *on* substate. Both schemes are reducible because there is only one substate in the *on* state. Even though they reflect very different mechanisms, it is possible to make $\phi_{on}(t)$ and $\phi_{off}(t)$ of the two schemes the same (e. g. by equating coefficients of the powers of the Laplace variable *s* of $\bar{\phi}_{on}(s)$ and $\bar{\phi}_{off}(s)$ from the two schemes ($\bar{g}(s) = \int_0^\infty g(t)e^{-st}dt$), which results in a set of equations relating the transition rates of the two models). The trajectories generated from the two schemes will be identical (in a statistical sense). The same is true for all three substate schemes (Fig. 2D - Fig. 2F), which are the simplest examples for reducible schemes. Recently, Witkoskie and Cao (2004) pointed out that counter to intuition schemes Fig. 2E, and Fig. 2F can be made indistinguishable using similarity transformation arguments. In the literature, in the context of enzyme kinetics, it has been suggested that it is possible to distinguish between schemes, Fig. 2B - Fig. 2C using more sophisticated analyses the trajectory (Edman and Rigler, 2000). This does not coincide with our findings here.

For irreducible kinetic schemes $\phi_{x,y}(t_1,t_2)$ is not factorized for at least one combination of $x,y = on, off$. In these cases, functions other than the waiting time PDFs contain additional information. Such functions are: (i) $\phi_{x,y}(t_1,t_2)$, $x,y = on, off$ itself (Lu et al., 1998 ; Cao, 2000; McManus et al., 1985; Colquhoun et al., 1996), as used in the pioneering work of Xie and collaborators (1998), and calculated for any kinetic scheme by Cao (2000); (ii) the *x-y* propagator for stationary processes, which is the probability density to be in state *y* at time *t* given that the process was in state *x* at time *0* (Lu et al, 1998 ; Edman et al., 1999; Edman and Rigler, 2000; Flomenbom et al.; Schenter et al., 1999 ; Boguñá et al., 2000). This determines the normalized state-correlation function, which is the bulk relaxation function; (iii) higher order state-propagators (Edman and Rigler, 2000; Schenter et al., 1999; Wang and Wolynes, 1995), or the corresponding higher order state-correlation functions; (iv) higher order successive waiting times PDFs, e. g. $\phi_{x,y,z}(t_1,t_2,t_3)$, $x,y,z = on, off$. Note that the functions in (i), (iii) & (iv) can be obtained only from single molecule experiments.

Which of these functions is the most useful in differentiating among irreducible schemes is still an open question. In practice, a function that involves many arguments will be noisy due to the limited number of events in the time-series. We have found that the PDF of the sum of (or, binned) successive waiting times, e. g. $\phi_{x+y}(t) = \int_0^\infty \int_0^\infty \delta(t - t_1 - t_2) \phi_{x,y}(t_1,t_2) dt_1 dt_2$, can not only be more accurately obtained from finite trajectories, but is more discriminatory than the equal successive waiting times PDF (Supp. Info.), e. g. $\phi_{x,y}(t,t)$ (Cao, 2000). $\phi_{x+y}(t)$ can be easily constructed from the trajectory by building the histogram of the random variable $t = t_1 + t_2$, obtained from *all* adjacent waiting times in the time-series. One can also calculate, in addition to the functions themselves, the difference between them and the product of the individual waiting time PDFs, e. g. $\Delta\phi_{x,y}(t_1,t_2) = \phi_{x,y}(t_1,t_2) - \phi_x(t_1)\phi_y(t_2)$, and $\Delta\phi_{x+y}(t) = \phi_{x+y}(t) - \phi_x * \phi_y$, where $\phi_x * \phi_y = \int_0^t \phi_x(t-\tau)\phi_y(\tau)d\tau$. These differences vanish for reducible schemes.

**DISCUSSION AND CONCLUDING REMARKS**

Given a two-state trajectory, after constructing $\phi_{on}(t)$ and $\phi_{off}(t)$, one should immediately determine whether the underlying kinetic scheme is reducible using Eq.

(1). Due to the finite length of the trajectory, the moments of $\phi_{x,y}(t_1,t_2)$ can be more accurately calculated than the PDF, and should be compared to the corresponding products of the moments of $\phi_x(t)$ and $\phi_y(t)$. Another test compares the bulk relaxation function (the state-correlation function) obtained directly from the trajectory, with the corresponding theoretical result for a TSSM process (Flomenbom et al.). The expression for the bulk relaxation function for a stationary TSSM is known, in Laplace space, for arbitrary waiting time PDFs (see equation 3.15 in Boguñá et al., 2000), so one can plug in the Laplace transforms of the experimental $\phi_{on}(t)$ and $\phi_{off}(t)$ into this expression, and invert the result, either analytically or numerically, back into the time domain. If the experimental bulk relaxation function and the theoretical one for a TSSM process with the experimental $\phi_{on}(t)$ and $\phi_{off}(t)$ coincide, the scheme is reducible, and no further analysis is required. Another simple and informative analysis method involves the trajectory of the waiting times as a function of the occurrence index. Correlations between waiting times can be detected more easily from this trajectory than the *on-off* trajectory, and used to learn about the scheme transition rates (Fig. 3).

To conclude, we note that some of the fundamental concepts presented in this work were already been used in the analyses of the catalytic activity of individual lipase molecules (Flomenbom et al.). In this case, the *off* waiting time PDF was best fitted to a stretched exponent. The bulk relaxation function test was then applied, and the kinetic scheme was shown to be irreducible. Additionally, clusters of fast events were detected in the ordered *off* waiting times trajectory (similar to Fig. 3A), indicating that single lipase molecules display correlations in their activity. These findings were combined to build a kinetic scheme that involves reaction and conformational changes simultaneously, and to extract some of the conformational and reaction rate values.

**FIGURE CAPTIONS**

Figure 1 *On-off* trajectories as a function of time. These trajectories were obtained by simulating the kinetic schemes shown in Fig. 2K (A) and Fig. 2L (B). The transition rate values are given in Fig. 3.

Figure 2 A set of schemes containing black circled *off* substates and red squared *on* substates, which can be used to produce *on-off* trajectories. A - F Reducible schemes with only reversible connections. G - I Reducible schemes with irreversible connections. J – TSSM model described only by the waiting time PDFs $\phi_{on}(t)$ and $\phi_{off}(t)$. K - The simplest irreducible model is a four-substate model. L – An example of a reducible four-substate model.

Figure 3 O*ff* waiting times trajectories as a function of the occurrence index corresponding to the *on-off* trajectories in Fig. 1, produced from the irreducible (Fig. 2K), and reducible (Fig. 2L) four-substate schemes. $\phi_{on}(t)$ and $\phi_{off}(t)$ for the two schemes are the same, by setting ($k_{ji}$ is the transition rate from substate $i$ to $j$), $k_{21} = 1$, $k_{12} = 0.09$, $k_{32} = 0.01$, $k_{23} = 0.1$, $k_{43} = 0.9$, & $k_{34} = 0.1$ for the irreducible one, and $k_{21} = 0.1818$, $k_{12} = 0.36818$, $k_{32} = 0.55$, $k_{23} = 0.495$, $k_{43} = 0.405$, & $k_{34} = 0.2$, for the reducible one, found by comparing $\bar{\phi}_{on}(s)$ and $\bar{\phi}_{off}(s)$ of the two models. In the ordered waiting times trajectory from the irreducible scheme similar waiting times tend to follow each other (A), where from the reducible one, the waiting times are randomly distributed (B). By applying a threshold on this trajectory, which separates the fast from the slow events, one can estimate the transition rates $k_{ji}$ by calculating the average of the fast and slow *off* waiting times, given by $\bar{t}_{off,fast} \approx 1/(k_{23} + k_{43})$ and $\bar{t}_{off,slow} \approx 1/(k_{12} + k_{32})$, and the average number of successive fast and slow *off* waiting times, given by, $\bar{n}_{off,fast} \approx k_{43}/k_{23}$ and $\bar{n}_{off,slow} \approx 2 + k_{12}/k_{32}$.

Flomenbom, O. Fig. 1

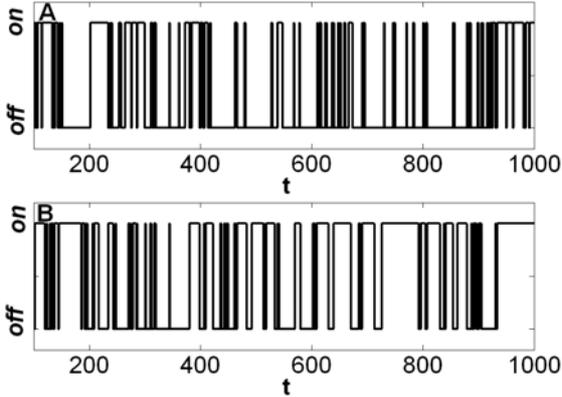

Flomenbom, O. Fig. 2

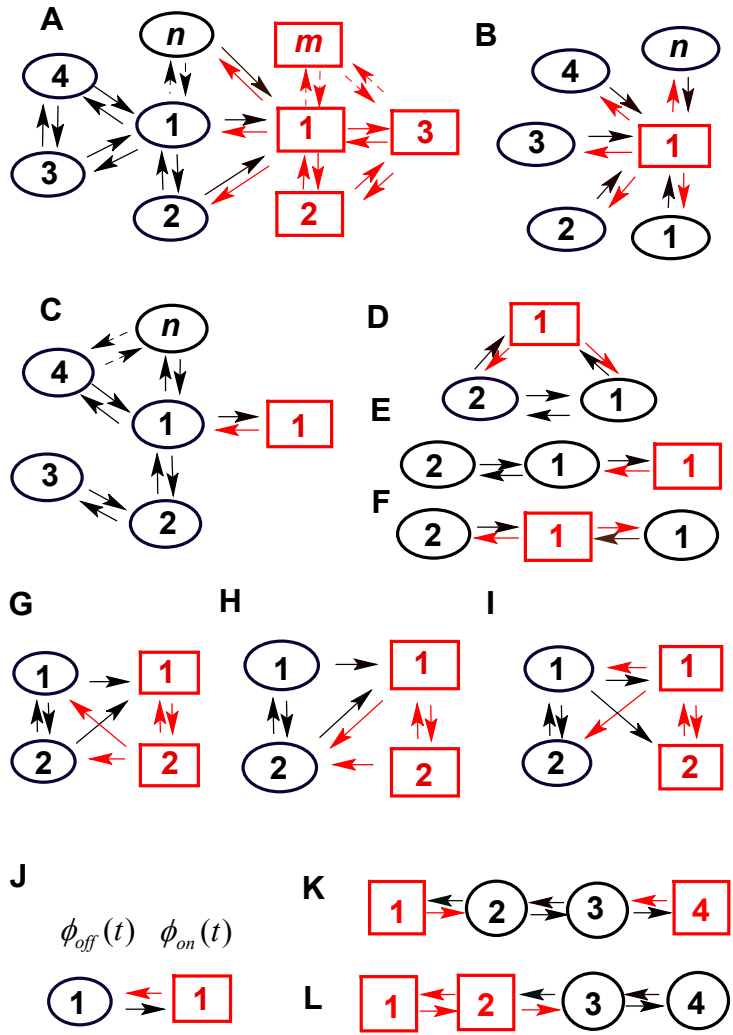

Flomenbom, O. Fig. 3

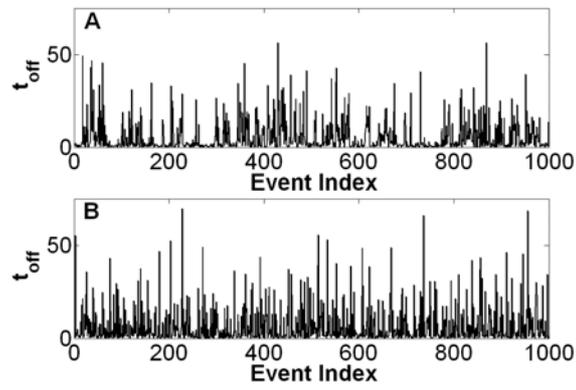

# Supplementary information

**What can one learn from two-state single molecule trajectories?**

Ophir Flomenbom, Joseph Klafter, and Attila Szabo

To demonstrate the advantage of $\phi_{x+y}(t)$ over $\phi_{x,y}(t) = \phi_{x,y}(t,t)$, we consider the simplest irreducible scheme, which is a four-substate scheme (Fig. 2k). The scheme is defined by a set transition rates $k_{ji}$ ($k_{ji}$ is the transition rate from substate $i$ to $j$), where substates 1 & 4 belong to the *on* state and substaes 2 & 3 belong to the *off* state. In what follows we set, $k_{21} = 1$, $k_{12} = pk$, $k_{32} = (1-p)k$, $k_{23} = 1-p$, $k_{43} = p$, $k_{34} = k$, and determine the effect of changing $p$ (= 0.9 & 0.3) and $k$ (= 0.5, 0.1, & 0.02) on the shape of $\Delta\phi_{off,off}(t)$, $\phi_{off,off}(t)$, $\phi_{off}^2(t)$, $\Delta\phi_{off+off}(t)$, $\phi_{off+off}(t)$, and $\phi_{off} * \phi_{off}$ (analytical expressions for these will be given elsewhere). Here, $k$ determines the asymmetry of the scheme, and for $k = 1$ the system is symmetric thus reducible, and both $\Delta\phi_{off,off}(t)$ and $\Delta\phi_{off+off}(t)$ consequently vanish.

$\Delta\phi_{off,off}(t)$, $\phi_{off,off}(t)$, and $\phi_{off}^2(t)$ possess the same shape for the parameter values examined (Figs. 1sA-1sD). The peak of $\Delta\phi_{off,off}(t)$ for larger times (Cao, 2000) is two orders of magnitude smaller than the maximal signal value (Figs. 1sC-1sF). $\phi_{off+off}(t)$, and $\phi_{off} * \phi_{off}$ are more sensitive to changes in the parameters (Figs. 2sA-2sB). For *p=0.9* as $k$ decreases a second peak emerges for $\phi_{off+off}(t)$ shown as a shoulder for $k = 0.1$ and as a small peak for $k = 0.02$ (the amplitude of the second peak as $p \to 1$ can be approximated by $k_{12}e^{-1}/2$). For *p=0.3*, where during each *off* event several transitions between substates 2 & 3 occur, both peaks are observable for a wide range of *k* values.

Thus, this example demonstrates that $\phi_{off+off}(t)$ and $\Delta\phi_{off+off}(t)$ (at least for some range of parameters as the examined ones) supply more information about the scheme details than $\phi_{off,off}(t)$ and $\Delta\phi_{off,off}(t)$.

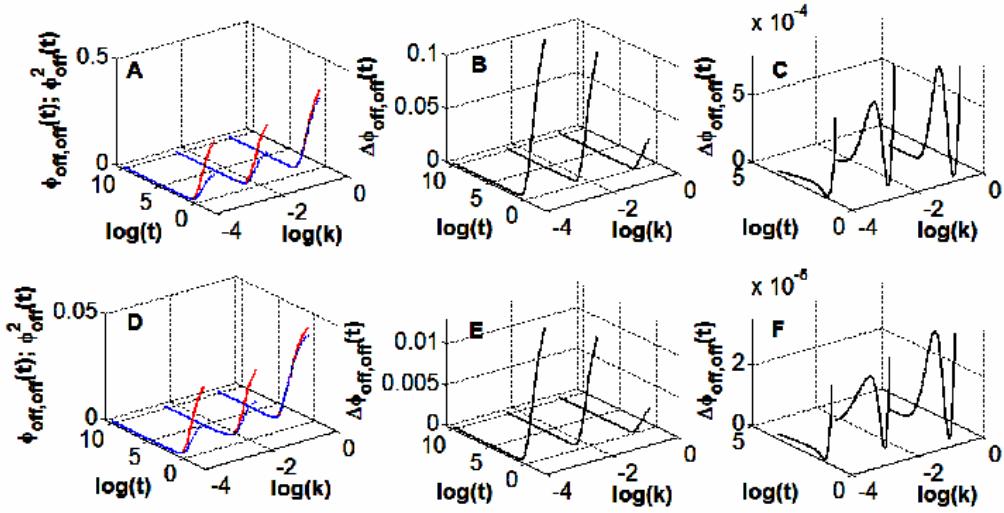

**Figure 1s A-F** $\phi_{off,off}(t)$, $\phi^2_{off}(t)$ (**A & D**), and $\Delta\phi_{off,off}(t)$ (**B-C & E-F**), for the four-substate irreducible scheme (Fig. 2K), with the (arbitrary units) parameters, $k_{21} = 1$, $k_{12} = pk$, $k_{32} = (1-p)k$, $k_{23} = 1-p$, $k_{43} = p$, $k_{34} = k$. In each plot, curves for three values of $k$ (= 0.5, 0.1, & 0.02) are shown, where for the upper plots $p = 0.9$, and for the lower plots $p = 0.3$. $\Delta\phi_{off,off}(t)$, $\phi_{off,off}(t)$, and $\phi^2_{off}(t)$ possess the same shape for the checked range of parameter values. The large time peak of $\Delta\phi_{off,off}(t)$ (**C & F**) is two orders of magnitude smaller than the maximal signal value.

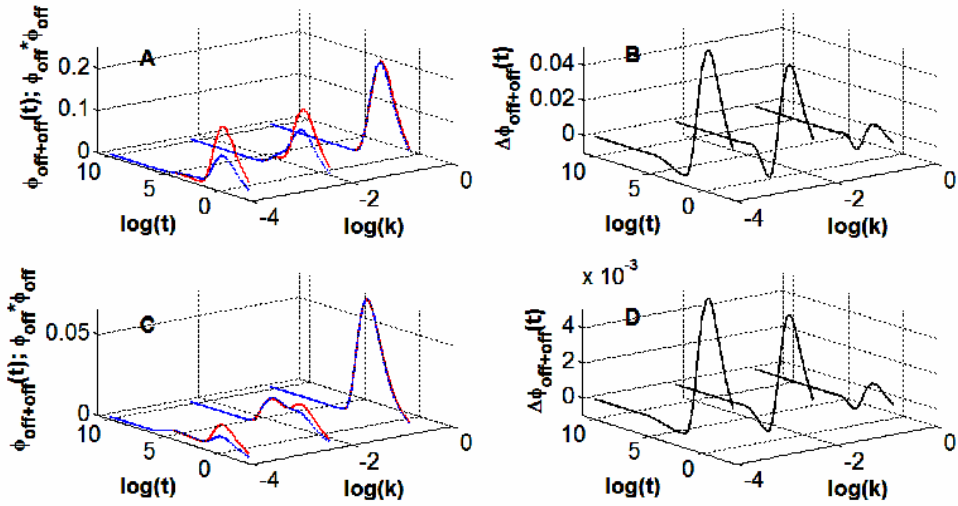

**Figure 2s A-D**

$\Delta\phi_{off+off}(t)$, $\phi_{off+off}(t)$, and $\phi_{off}*\phi_{off}$ are shown for the same range of parameters as in Fig. 1s. Upper plots ($p=0.9$): for $k=0.5$ the system is close of being symmetric, so that $\phi_{off+off}(t)$ and $\phi_{off}*\phi_{off}$ are similar (**A**), namely, $\Delta\phi_{off+off}(t)$ is (relatively to the amplitude of the functions themselves) small (**B**). As $k$ decreases a second peak appears in $\phi_{off+off}(t)$ representing (*i*) the two very different timescales of $\phi_{off}(t)$, and (*ii*) the fact that events with similar waiting time are clustered. The peak appears as a shoulder for $k=0.1$, and as a small peak for $k=0.02$ (**A**), where its amplitude as $p\to 1$ can be approximated by $k_{12}e^{-1}/2$. As $k$ decreases, $\Delta\phi_{off+off}(t)$ amplitude increases, although its basic shape retains. Lower plots ($p=0.3$): Here, the amplitudes of the two peaks are comparable (**C**), which is a signature that each *off* event consists of several transitions between substates 2 & 3 occur before leaving to the *on* state. $\Delta\phi_{off+off}(t)$ shows the same general behavior with $k$ as for the higher value of $p$ (**D**).